# Nanomodular Electronics

Microelectronics made anywhere,
anytime, by anyone

Michael Filler
*Georgia Institute of Technology*

Ben Reinhardt
*Speculative Technologies*

## Abstract

It may be possible to reinvent how microelectronics are made using a two step process:

1. Synthesizing modular, nanometer-scale components – transistors, sensors, and other devices – and suspending them in a liquid "ink" for storage or transport.

2. Using a 3D-printer-like machine to create circuits by placing and wiring the components.

These *nanomodular electronics* could enable a "fab in a box" and make fabricating microelectronics as straightforward as printing this document.

The process of using light and chemicals to construct microelectronics from a single piece of silicon has both enabled Moore's Law and created many unfortunate downstream effects: industry consolidation in volatile parts of the world, fragile supply chains, high overhead for making custom circuits, and large barriers for innovating on the process because everything is so tightly coupled. Incremental improvements to the current system are possible but won't get around the fundamental limitations of the underlying process paradigm – the planar process.

Even metrics long synonymous with the progress of planar processing can be deceptive. The transistors at the core of microelectronics, for example, while cheaper than they once were, are still shockingly expensive compared to other manufactured goods. At roughly $10 billion per kilogram, transistors cost thousands to millions of times more than even drugs, which are created in bulk chemical processes that scale with

volume. If you could create transistors and other circuit components in the same way, making microelectronics could be as ubiquitous as writing software. However, changing how we make transistors requires an entirely new *process* for creating microelectronics.

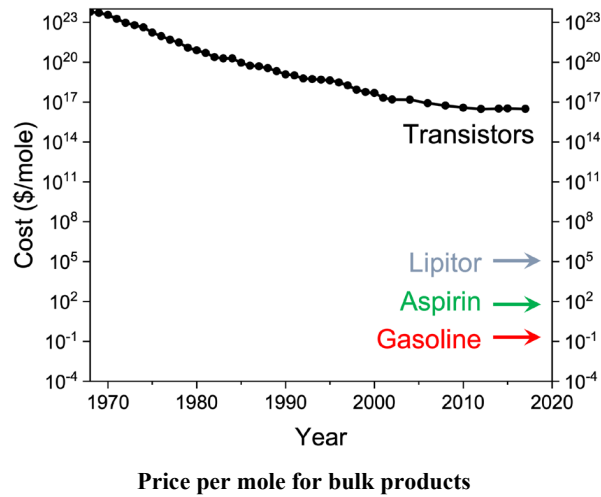

**Price per mole for bulk products**

Developments in *nanotechnology, colloidal chemistry, precision additive manufacturing*, and *computer vision* suggest that this new process is possible!

Creating nanomodular electronics needs a research program that coordinates several parallel component-focused projects towards a single goal: a student competition to explore the possibility space of nanomodular electronics and uncover compelling uses.

A student competition is a powerful "forcing function" for creating a general purpose technology. It ensures the system will be flexible and consistent enough to accommodate unexpected ideas and non-expert users. Both synthetic biology and VLSI electronics became widespread because immature tools were put in the hands of creative and energetic students.

If the history of general purpose technologies is a guide, nanomodular electronics could have impact far beyond their obvious near term applications like tamper-resistant electronics with unique identifiers, physically implemented neural networks, and myriad low-volume applications that depend on custom circuits.

# Table of Contents

# I. What are we trying to do?

## A. Create a new process for making microelectronics

The big goal is to unlock a new process for making microelectronics:

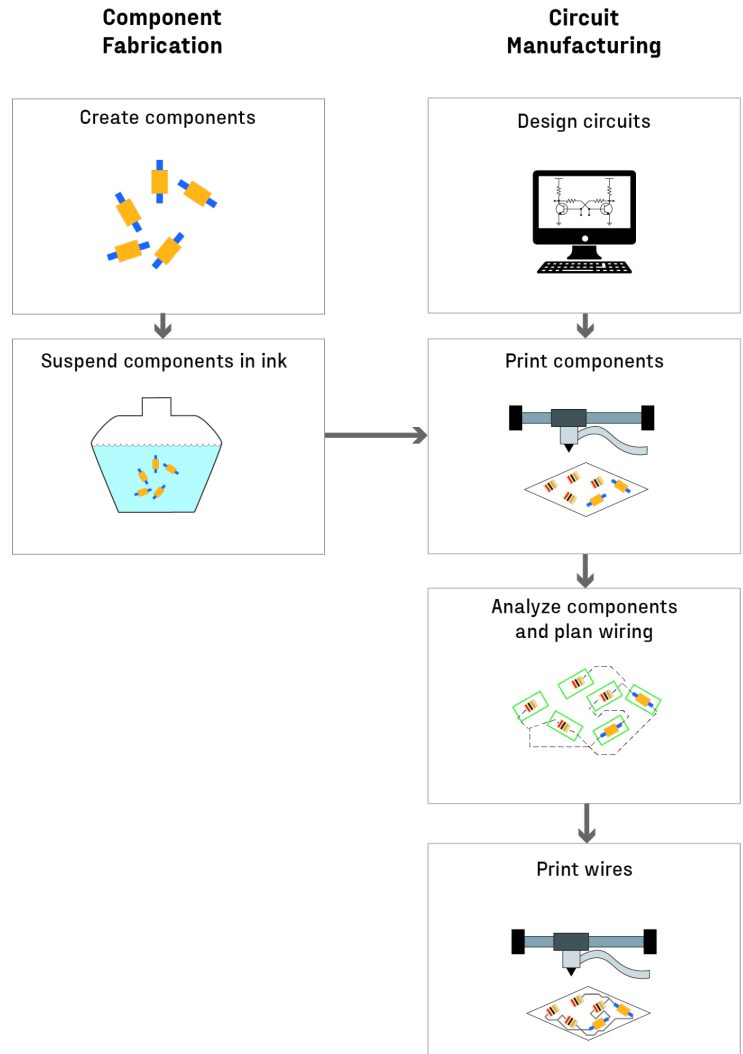

**Figure 1: A diagram of the nanomodular electronics process.**

### 1. Component fabrication

**Create components.** The process starts by creating modular, nanometer-scale components – transistors, sensors, etc. The program will bootstrap off existing microelectronics manufacturing by making components via photolithography and lifting them off their wafer with components made in bulk with "bottom-up" chemical processes in the long run.

**Suspend components in ink.** A colloidal "ink" contains the components in a separated, uncontaminated, and stable status until deposition.

## 2. Circuit manufacturing

**Design circuits.** Software generates a circuit layout based on a desired circuit function.

**Print components.** A 3D-printer-like tool deposits components from one or more inks.

**Analyze components and plan wiring.** A vision system detects the orientation and position of the components (which will have some stochasticity), plans a wiring path among them to create the intended circuit.

**Print wires.** A 3D-printer-like tool lays down conductive wires among the components based on the wiring plan.

Factoring [1] component and circuit fabrication creates a process similar to that of making printed circuit boards, albeit orders of magnitude smaller ($10^{-3}$ m for printed circuit boards versus $10^{-8}$ m for microelectronics), leading to an entirely different set of technology requirements and possibilities.

Today, the process of creating microelectronics occurs in a centralized location, on a single substrate, with no intermediate products between raw silicon wafers and completed circuits. By factoring the process, making microelectronics could be as straightforward as printing this document.

## B. Program goal: Create a manufacturing system to support the Nanomodular Electronics Challenge

We considered three potential goals for the program at the Workshop on On-Demand Integrated Circuits held from March 23-25, 2022 in Miami, FL:

- Build a common circuit (e.g., a [555 timer](#)) to establish that nanomodular electronics can make circuits that people already use.

- Fabricate cryptographic circuits that take advantage of nanomodular electronics' mass-customization capabilities.

- Create a manufacturing system that supports a student competition to explore the creative applications of nanomodular electronics.

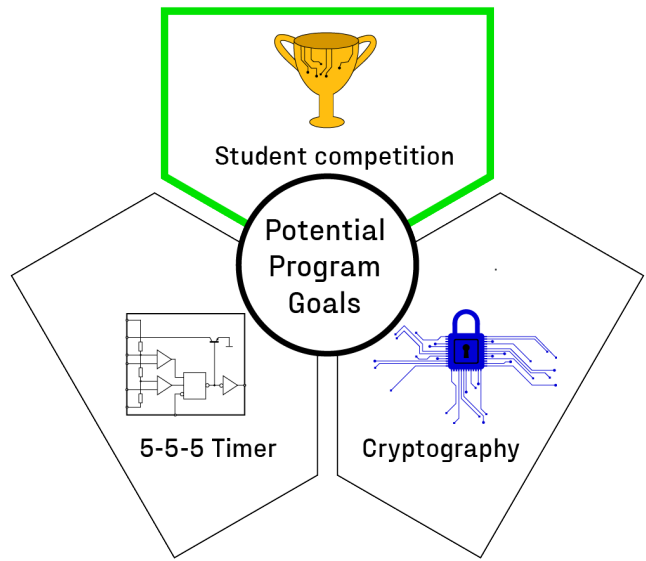

**Figure 2: A manufacturing system that supports a student competition is the best program goal.**

Building a manufacturing system to enable and support a student competition – the *Nanomodular Electronics Challenge* – could create a nourishing environment for a new general purpose technology (GPT):

- Students and competitions have had a significant impact on the development of several modern GPTs.

- Student competitions create both a community and a serious context of use.

- Competitions push against premature overspecialization.

- Competitions help uncover unexpected applications.

Let's walk through each of these benefits:

**Students and competitions have had a significant impact on the development of several modern GPTs.** In the International Genetically Engineered Machine (iGEM) competition, multidisciplinary student teams from around the world build projects with a toolkit of cutting-edge synthetic biology "parts." It has grown from six teams to more than 3600 over the past two decades. Teams mix and match biological parts to create systems that accomplish self-selected goals which are judged on a [number of criteria](). iGEM's success is obvious: it engendered tremendous interest in synthetic biology and led to an explosion in uses.

Carver Mead and Lynn Conway's project course unlocked large-scale chip design by enabling students to use cutting-edge very large-scale integration (VLSI) design tools to create their own chips. The course exploded on the young [ARPAnet](), leading to better systems and practices around VLSI tools and contributing to the creation of Silicon Graphics, which in turn left an indelible mark on the semiconductor industry.

A nanomodular electronics program would build a system that serves the same function – a platform students could use to springboard creative research projects.

**Student competitions create both a community and a serious context of use.** Successful technologies all have people who care about using them for something they care about – a serious context of use. When people care deeply about a context of use enabled by a new technology, they are motivated to provide the creators with discernible, useful, and timely feedback, often about things the creators would otherwise not think important.

It is important to note that people often care deeply about contexts of use that others would not think of as "useful": sports are often an early market for new technologies, like carbon fiber golf clubs, because people care deeply about their hobbies.

Competitions are a way of creating an artificial, but serious context of use. If successful, a student competition would bootstrap a community of people who take tinkering with nanomodular electronics seriously, creating an honest feedback loop for improving and advancing the technology.

**Competitions push against premature overspecialization.** Building a competition-worthy platform is a strong forcing function for GPTs: a tool with enough flexibility to accommodate unexpected ideas while being consistent enough for non-experts to use.

New technologies can fail to become GPTs because they are only suited for a particular application, making them unusable for other purposes. For example, if transistors had been used exclusively as amplifiers in undersea cables as originally envisioned, it is unclear whether modern computing would have emerged.

A student competition, where it is impossible to predict what the competitors will ask of the system, demands generality.

**Competitions help uncover unexpected applications.** The "killer applications" of a GPT are difficult to pinpoint in its early development. Therefore, quick and continuous exploration of the potential applications for nanomodular electronics is essential and can be achieved through a student competition.

Sufficiently motivated students have more time than busy professionals, fresh minds and energy, increasing the likelihood of discovering important contexts of use.

Recurring competitions also take advantage of the incremental improvements to the system and can continuously reveal new applications.

**A student competition has other advantages.** If the goal is "do cool stuff," there is little direct comparison between nanomodular electronics and traditional microelectronics, giving it time to grow. 3D printing followed a similar trajectory in the maker community long before it had commercial applications.

Modular systems coupled with play, e.g., Minecraft, Roblox, or Lego, have historically led to far more impressive outcomes than envisioned. Coupled with the capability to be more than a toy, there is reason to be optimistic about similar outcomes for nanomodular electronics.

## C. Technical benchmarks for the Nanomodular Electronics Challenge

The Nanomodular Electronics Challenge requires good, not exceptional, performance, enough consistency not to frustrate competitors, and enough flexibility to encourage exploration of the affordances of nanomodular electronics. Thus, the program will target a system capable of:

- Late-1990s and early-2000s performance, *e.g.,* the **180 nm process node** [2]. Component fabrication is easy relative to more advanced nodes, as the 180 nm node was one of the last where photolithography did not require "tricks" to achieve features smaller than the wavelength of light. Yet, the transistors are fast enough for long-distance wireless communication and create a large design space for competitors.

- Circuits with ~**10,000 components** to enable a vast array of circuit functions – from small microprocessors and memory arrays to smart stickers and physically unclonable cryptographic keys – while avoiding the need for obscene yields or highly optimized routing algorithms. For comparison, the transistor counts of the famous Intel 4004 and MOS Technology 6502 are 2,300 and 3,510, respectively.

- **Wire printing speeds** greater than ~1 mm per second could interconnect ~10,000 components in under 10 minutes, allowing students to test and iterate on designs in real-time.

## D. The importance of a coordinated research program

Nanomodular electronics is a *systems* challenge. The process requires subsystems for creating nanoscale transistors and other components in bulk, storing and transporting them, placing them in specified locations, and connecting them while dealing with the stochasticity inherent in nanoscale systems.

There are many dependencies among these subsystems, a few being:

- The width of interconnect wires and their printing precision must be less than the dimensions of the smallest component features.

- Components must be transportable while not increasing contact resistance between the component and circuit.

- The wiring printer must accommodate some amount of randomness in component placement as it lays down wires among them.

Navigating those tradeoffs requires tight coordination between the groups working on each component and iterating on a complete system as early in the process as possible. Building a functional *system* acts as a forcing function for coordination – no individual project can be considered successful unless the entire system hits its goals. Projects within the program will need to adjust based on discoveries or setbacks in other projects. This *systems research* is crucial to the creation of new general-purpose technologies and manufacturing processes.

It's worth noting that this program doesn't have a clear institutional home in academia, industry, or government. The ratio of grungy engineering to novel insights is too low for academia. At the same time, it is still speculative research that would make a terrible business proposition until the technology is more advanced. Nanomodular electronics is too far outside of the government's [Overton window](#) to merit support from organizations like DARPA or legislation like the CHIPS Act.

We hope this roadmap will unlock work that drives technology to a point where some of these institutional constraints no longer apply.

# II. How are microelectronics made today? What are the limitations of the current system?

## A. The basics of traditional microelectronics manufacturing

Let's start with some explain-it-like-I'm-five preliminaries that those familiar with semiconductor manufacturing can skip:

- Integrated circuits are made of many semiconductor *components* connected by tiny electrical wires to create *circuits*. The primary component that people talk about in computing is the transistor. Other useful components include sensors, resistors, capacitors, etc.

- The *planar process* creates integrated circuits on a semiconductor *wafer* in a layer-by-layer fashion using a variety of additive (chemical vapor deposition) and subtractive (photolithographic patterning and chemical etching) steps. Transistors were originally embedded in the silicon wafer leaving the top surface flat, hence the name "planar" process. Modern versions of the planar process do not embed transistors in the wafer but instead position them on top of it.

- Semiconductor wafers are almost all made of silicon. Older nodes use 200 mm diameter wafers and newer nodes (after ~2000) use 300 mm diameter wafers. Each wafer can hold one or more, and usually many, integrated circuits.

- After wafer manufacturing, the integrated circuits are cut out into individual *chips* and packaged.

- The planar process creates *monolithic* integrated circuits. All the components and wires are created in a single, unbroken process ([up to 1400](#) steps!) that starts with raw silicon wafers and ends with complete integrated circuits. Individual components of the circuits are inseparable. If one step goes awry, the entire wafer can be scrap.

- The *node size* refers approximately to the smallest device feature (like the channel width of a transistor) that a specific process can create. As node sizes get smaller, transistors become denser, allowing more transistors per chip and per wafer. Moore's Law refers to the exponential growth in transistors per chip. As of 2022, the 5 nm process is the smallest node in production. However, many larger nodes remain in production. Arduinos still use chips made using the 350 nm node.

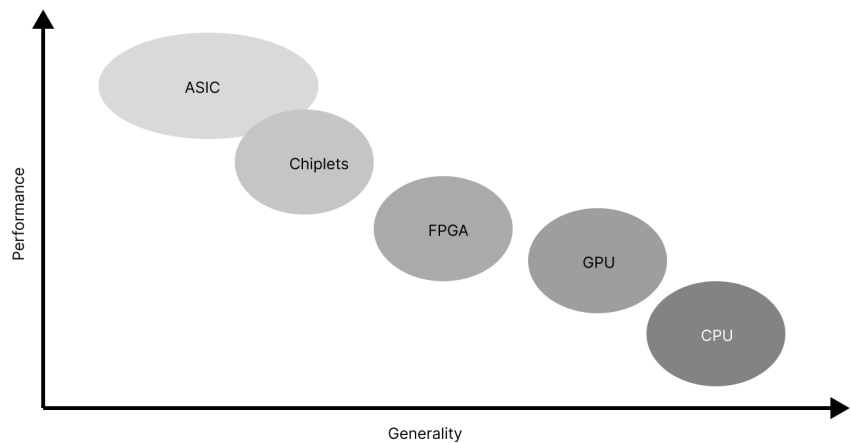

**Figure 3: Comparison of trade-offs between performance and generality in traditional microelectronics.**

The planar process can create chips that occupy a large swath of a design space that trade-off generality for performance. In addition to the workhorse circuits for general-purpose computing, application specific integrated circuits (ASICs) are specialized chips for specific functions, from computer vision to matrix multiplication. Field programmable gate arrays (FPGAs) are ASICs that can be reprogrammed to simulate one specific circuit at any given time. FPGAs enable in-the-field customization by taking a performance hit compared to a single-function ASIC but can be much more performant than a general-purpose circuit for any specific task.

There are several core numbers at play in the economics of microelectronics manufacturing:

- The capital expenses (capex) of creating the production line. These are one-off costs of building the building, buying the machinery, training people to operate it, and (this one often gets left out) figuring out how to get everything working.

- The recurring costs to the manufacturer — these occur per *wafer* and include consumables like ultra-pure water, chemicals, and electricity as well as labor. (Also known as operational expenses or opex.)

- The non-recurring (i.e., one-off) costs associated with creating a new chip — design, creating the masks, testing them, and iterating.

- The price that the manufacturer can charge for a wafer.

- The price that a retailer can charge for a chip.

There are several metrics that change monotonically as node size decreases:

- Capex increases. A new 180 nm fab costs ~$100M; a new 5 nm fab costs ~$10B.

- Opex increases. By some calculations, a 90 nm wafer costs $411 to manufacture while a 5 nm wafer costs $4235 to manufacture [3].

- Chip design cost increases. 5 nm chips are about 20 times more expensive to design than 28 nm chips [4].

- Transistor density increases. A 90 nm wafer has 1.45 million transistors per square millimeter while a 5 nm wafer has roughly 1.8 billion transistors per square millimeter.

- The power per computation that each chip uses while operating decreases.

The majority of the *profit* in the chip market comes from the leading edge chips with the highest transistor densities. Customers who value power efficiency and computations per second like militaries and data centers are willing to pay a premium for these chips. This demand makes the margins (and thus profit) for manufacturing more advanced nodes significantly larger than for legacy nodes because the sale price per wafer increases even faster than increasing capex and opex.

However, older nodes account for a large fraction of *sales* in the chip market. Processes older than the 60 nm node account for 90% of the market and 23% of the market is processes older than the 180 nm node. Car manufacturers are now a huge consumer of chips and tend to lock in at older, cheaper nodes. The same goes for many appliances and IoT devices. As a result, almost all machines for making older nodes are still in use. Used machines are expensive because the market for them is illiquid — people don't often sell machines for older nodes and the continued demand for older node chips means the demand for these machines lasts for decades.

## B. The constraints of traditional microelectronics manufacturing

It is easy to see the cost, complexity, and sensitivity of traditional microelectronics manufacturing as well as the broader challenges of the industry as isolated issues. However, each seemingly independent constraint is downstream of the foundational assumptions of the planar process. Seventy years of continuous improvement has created an exquisite, hyper-optimized system.

### 1. Technical constraints

Today's circuits are monolithic - components are built layer by layer, material by material, on a single semiconductor wafer. The intimate connection between the components and wafer makes it impossible to remove or replace components – a defective component can ruin the entire circuit. As a result, yields need to be close to 100%. Near-perfection requires pristine, spatially uniform processing environments and purified consumables. These purity demands drive up costs and cycle time. Microelectronics manufacturing is unusual: most manufacturing processes are not bottlenecked by yield in this way, suggesting that there may be other approaches.

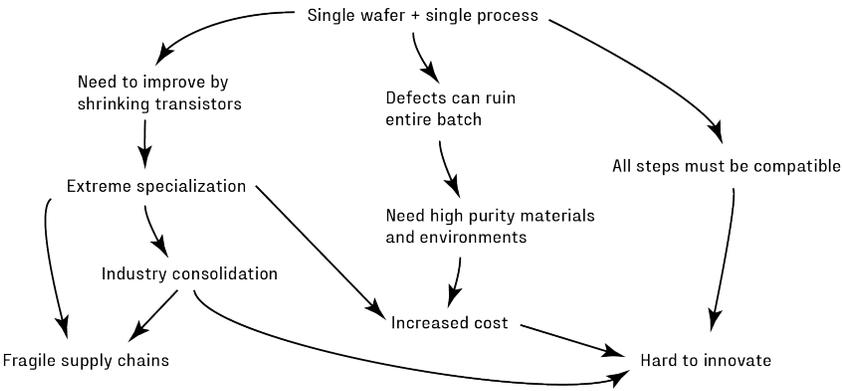

**Figure 4: Many constraints on how we make microelectronics today are downstream of the planar process**

Monolithic circuit fabrication has no natural breakpoints. Semiconductor wafers and raw materials enter the fab and only complete circuits exit. (The equivalent would be a car factory that takes raw metal, glass, plastic, etc. as inputs and outputs a fully functioning car.) As a result, all steps must be compatible with all others, restricting the space of possible processes, conditions, and locations where different components can be fabricated. The sheer number of steps to get to the final circuit limits fabrication speed, no matter how fast the individual steps.

The planar process is optimized around a small set of materials. Silicon is usually a given, with subsequent steps all tuned to its specific properties. It is possible to make components out of other materials: germanium, gallium arsenide, gallium nitride, silicon carbide, and even organics. While there were good historical reasons to initially focus on silicon (operational stability, cost, simplicity, and useful temperature range), the planar process cannot unlock the benefits of multiple materials on a single chip without reinventing vast swaths of the process.

## 2. Economic constraints

Firm specialization has driven manufacturing progress, but also created several constraints. Chip manufacturers depend on thousands of companies across several continents to provide the machines, consumables, and services necessary for consistent near-perfect results. None of the large manufacturers are vertically integrated. (A large amount of this coordination happens today via the IEEE International

Roadmap for Devices and Systems (IRDS).) There are several specialized suppliers, like ASML, that create single points of failure. While a plethora of specialized firms has economic advantages, they make the entire semiconductor industry slow to respond to changing conditions. Additionally, few firms have the resources or scope to make changes across the entire process.

Complex supply chains with many entities handling both chips and tools also mean there are numerous points where the security of a chip can be compromised, from counterfeiting to hardware-level vulnerabilities.

Surprisingly, the chip shortages induced by the COVID-19 pandemic are primarily in older nodes, not leading-edge chips. Manufacturers didn't respond to increased demand by spinning up new fabs and increasing supply because of a combination of three factors: the economics of older nodes, the time it takes to build a fab, and the cyclicality of the market for older nodes. It takes years to spin up even an older node fab. The market for older nodes has historically fluctuated, so by the time the fab is online, many manufacturers predict that demand will have died down. Even older fabs are expensive enough that they need to be running near full capacity to recoup the cost of building them, so manufacturers believe that building new older-node capacity in response to immediate market conditions is a bad idea. By contrast, they see a monotonically increasing demand for bleeding edge nodes, as long as the bleeding edge goalpost keeps moving — hence the massive amount of R&D and new fab dollars.

Building and operating a semiconductor fab requires deep technical experience. There is a massive amount of tacit knowledge needed, even for older nodes. Machines are not plug and play: every fab has new, unique settings and processes, requiring trial-and-error and hands-on experience to ensure consistent, reliable results. There are fewer individuals with the requisite knowledge due to industry consolidation and an aging workforce. On top of that, since each manufacturer must construct all circuit components, they must have expertise in all process steps.

# III. What is technically new in your approach? Why do you think it will be successful?

The technical difference between traditional planar processing and the nanomodular electronics process boils down to factoring [1] microelectronics creation into two discrete steps: component fabrication and circuit manufacturing.

## A. Factoring component fabrication and circuit manufacturing creates myriad downstream technical differences

### 1. Component fabrication

**Components are modular.** As "nanomodular electronics" implies, a key technical difference from traditional electronics is that components are modular: they can be mixed and matched with other components in arbitrary combinations after they are manufactured. Modular components need stable interfaces like contacts for external wiring. They also need all the elements for their operation in a self-contained package. In the case of a transistor, that means the source, channel, drain, gate stack, and contacts.

**Components with a wider range of materials, structures, physics, and specifications.** Modular components make it much easier to incorporate a wide range of materials in the same circuit. Some potential examples include:

- III-V materials (e.g. gallium arsenide) that are faster than silicon, more resistant to radiation, and important to solid-state light sources and high-speed photodetectors.

- Wide band gap semiconductors (e.g., gallium nitride, silicon carbide) that are useful in high power applications, high temperature environments, and ultraviolet photodetection.

- Narrow band gap semiconductors (e.g., germanium, mercury-cadmium-telluride) are valuable for infrared spectroscopy and thermal detection.

- Components made from materials that are not semiconductors, such as magnetic materials in active components, metal oxides in resistors, or polymer dielectrics in capacitors. (In planar processing, passives are usually composed of semiconductors or eliminated altogether.)

- Components emerging from R&D labs made from entirely new materials or based on new physics.

- Components that work best at cryogenic temperatures or are radiation resistant for use in space.

Ultimately, circuit makers should be able to choose components from a catalog based on their function, specifications, and pricing: a Digi-Key catalog for individual nanoscale components.

**Chemicals-like manufacturing of components.** Non-monolithic components don't need a substrate. Substrate-independence opens the door to components made via bottom-up processes. These components, in turn, enable chemistry-like manufacturing, which scales with volume instead of surface area.

**Components packaged in new ways.** Component packaging could be selected based on need. Packaging designed for implants could prevent operational degradation in biological environments and ensure biocompatibility, whereas packaging designed for satellites could tolerate ionizing radiation and low temperatures.

## 2. Circuit creation

**Low temperature and non-toxic circuit creation.** While creating high-performance components requires temperatures from 400-700 ºC and hazardous chemicals like hydrofluoric acid, creating metal wires and insulators is possible at temperatures below 200 ºC with fewer, less hazardous chemicals. Separating the two steps means that high temperatures and toxic chemicals will not be needed at the point of circuit manufacturing. It also opens the door to making circuits on glass, plastic, paper, or even biological substrates.

**Printed wiring** simplifies tooling by making photolithographic mask sets for each circuit design unnecessary, removing a key contributor to cost and cycle time in traditional manufacturing.

**Additional types of components do not add complexity.** While each type of component requires its own set of processing steps, the complexity of the end-user tool for nanomodular electronics would remain relatively constant with additional types of components.

**Decoupled transistor size, spacing, and circuit speed.** There is a technical tradeoff in traditional microelectronics: signals can move faster in larger, further spaced wires but maximizing wafer usage demands more tightly packed transistors and wire density. High wire density causes neighboring wires to interact, resulting in unwanted signal propagation ("RC") delays. Circuit speed has been limited by these RC delays rather than transistor switching speeds since the 180 nm node [5].

Nanomodular electronics could break this tradeoff. Placing high performance transistors on inexpensive substrates relaxes the necessary transistor density. More widely spaced transistors would allow for larger, less dense wires, thus reducing delay. These lower-density components could create circuits that are faster than their traditional counterparts, limited only by transistor switching speed. As a bonus, lower transistor and wiring densities could permit faster heat removal or increase circuit flexibility.

**Routing flexibility.** Relaxed density requirements mean that not all deposited components must be routed, giving a routing algorithm much needed flexibility. For a given circuit design and component deposition process, algorithms could wire components with positions that maximize circuit performance and leave poorly placed or defective components unconnected.

3. Design

**Mixed-and-matched components.** Nanomodular electronics could heterogeneously integrate different materials and device types on the sub-micrometer scale, which is incredibly difficult with traditional microelectronics. Sub-micrometer heterogeneous integration could enable entirely new circuit functions and allow designers to engineer specific functions by physically co-locating components made of different materials.

**New form factors.** Nanomodular electronics could have topologies, mechanical properties, and densities that are hard for traditional electronics. Today's microelectronics are built on a rigid, flat silicon substrate. Components that are deposited instead of patterned and etched on a surface could be placed with a full six degrees of freedom, enabling more components per unit volume and naturally non-planar topologies.

Deposited (or embedded) components also enable nanomodular electronics to have mechanical properties that are hard to achieve on a semiconductor wafer: less brittle, stiff, and less likely to fracture.

**Interfacing components on different size scales.** Printed circuit boards combine discrete macroscopic components and integrated circuits. Nanomodularity breaks down this long-standing dichotomy, allowing electronic components big and small to be intimately combined in new ways.

4. Manufacturing process

**Distributed manufacturing.** Unlike traditional microelectronics production, with each circuit going from raw material to finished product in a single fab, in nanomodular electronics, components could be fabricated in one place with circuits created elsewhere. With reasonable specifications, the two steps could be done with minimal coordination in entirely different cities or countries.

**Component stockpiling.** The clean break between component fabrication and circuit creation in the two-step nanomodular process means that nanomodular components could be stored, perhaps as powders or inks. Stored components could be integrated into many different circuits, months, or years later.

**Rapid iteration.** The stability of the interface between components and circuits, once established, means that each can be iterated upon independently. New innovations could be "dropped in" more easily.

**Circuit extensibility.** The functionality of traditional electronics is permanently locked in at manufacture. By contrast, modular interfaces and discrete wiring processes could be amended or extended, either by the original manufacturer or someone else.

**Lower energy manufacturing**. Making transistors requires an energy input on the order of a billion megajoules per kilogram, much larger than the thousands of megajoules per kilogram needed for most pharmaceuticals. This energy difference suggests that transistors manufactured more like chemicals could drastically reduce the energy demands of microelectronics manufacturing.

**Additional process innovations.** Removing the constraints of a monolithic process should make it easier to change the process in the future like further factoring of steps (see fig. 5).

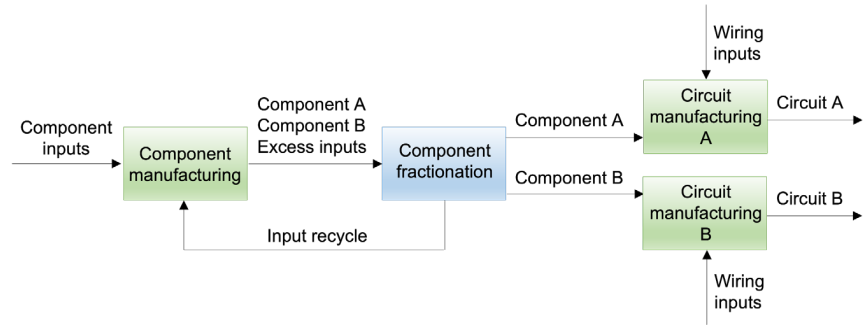

Figure 5: Example factoring of component manufacturing into a synthesis and fractionation step. Good components (component A) are sent to one circuit manufacturing process while better components (component B) are sent to a second circuit manufacturing process. The process recycles unused components to the component manufacturing step.

## B. Nanomodular electronics are not chiplets, open source silicon, printed electronics, nor nanotechnology

It might not be clear how building nanomodular electronics is different from current research towards improving electronics like chiplets, open source silicon, printed electronics, or nanotechnology more broadly. There are obvious similarities: chiplets focus on modularity; open source silicon aims to democratize microelectronics design; printed electronics use 3D printing techniques and inks made of nanoscale materials to build circuits; nanotechnology is concerned with functional nanoscale objects.

However, nanomodular electronics optimize for customization, cost, and extensibility. These other research programs optimize for other things, creating differences that will amplify over time:

- **Chiplets** optimize for performance by taking advantage of the existing planar process which also subjects them to the planar process' limitations on per-transistor cost, mass-customizability, and equipment requirements.

- **Open source silicon** uses open source design flows but assumes planar processing, meaning that fabrication still requires months and must currently rely on sponsor companies to cover the cost of manufacturing [6].

- **Printed electronics** are still monolithic: they are created whole cloth with no natural break-points. As a result, they require a variety of post-print processing steps so end-users still need a room full of chemicals. Because of the influence of each process step on many others, it will be hard to innovate on the process.

- **Nanotechnology** is certainly a component of nanomodular electronics, but it has little to do with building a functional manufacturing process that interfaces between many length-scales.

## C. Most components of a nanomodular electronics process exist and the system has no fundamental physical barriers

There are multiple reasons to be bullish on nanomodular electronics:

- Factoring manufacturing processes has consistently unlocked new pathways for continuous improvement [1].

- Physics does not rule out nanomodular electronics with performance comparable to traditional microelectronics.

- Researchers have demonstrated aspects of all the capabilities needed for nanomodular electronics in isolation.

### 1. Colloidal inks are ideal for storage, transport, and printing

Colloidal solutions consist of individual solid nanoparticles dispersed in a liquid that keeps the particles from clumping together or falling out of solution. A variety of nanoscale objects – nanocrystals, nanorods, and 2-D sheets composed of a variety of relevant materials – can be made into functional, stable colloids [7-9]. These approaches are increasingly viable for nanoparticles with complex internal structures [10].

Groups have also incorporated component inks into functional electronics systems. For example, using an ink of carbon nanotubes to fabricate logic gates on a 16-bit microprocessor [11]. While this work is a proof-point that semiconducting materials in inks can create high performance circuitry, there is still work to be done for two major reasons. First, except for using carbon nanotubes instead of silicon, creating the rest of the transistor as well as the wires used conventional fabs and processes. Second, nanotubes are far simpler than the transistors that a nanomodular electronics system will need, so adapting the ink isn't strictly a matter of swapping out nanotubes for nanomodular transistors.

## 2. Methods are available for controlling the placement of nanomodular components or accounting for placement imperfection

There are several approaches for placing nanomodular components that have trade-offs among precision, speed, and simplicity. Drop casting, spray coating, blade coating, and related techniques are relatively quick and easy but only provide control of *average* object density and alignment [12, 13].

More precise assembly techniques exist [14], but are slower and require specialized equipment. For instance, a combination of electron beam lithography patterning and DNA self-assembly allows complex patterns of metallic nanocrystals to be assembled with yields exceeding 95% [15]. Dielectrophoresis can assemble thousands of semiconductor nanowires with greater than 98.5% yield [16].

In either case, a nanomodular electronics manufacturing system could account for imperfect component placement. By combining imaging and machine learning, the system could classify as-placed nanomodular components [17-19] and route the desired circuit "on-the-fly" [20].

## 3. Emerging additive manufacturing methods have sufficient resolutions and rates

Additive manufacturing techniques such as electrohydrodynamic jet (e-jet) printing [21] and fountain pen nanolithography [22] offer the resolutions and rates required to wire nanomodular electronic circuits. E-jet printing uses an electric bias to pull ink from a small nozzle creating lines at rates on the order of 1-100 mm/sec with widths as small as 50 nm with current technology [23]. Fountain pen nanolithography uses capillary forces, sometimes coupled with oscillation, to extrude inks at similar rates and resolutions. As a bonus, the equipment fits in an area no larger than a desktop [24].

As a rule of thumb, the minimum wire width must be smaller than the component feature to be wired. In early nanomodular electronics systems, that feature will be the 180 nm gate of a transistor. Assuming a printing rate of 1 mm/sec, a standard component fan-out of 5, and an average (albeit large) component spacing of 10 um, more than 10,000 components could be wired in under 10 minutes.

Commercial additive manufacturing tools from companies like [Nano Dimension](#) are getting to the stage where printing passive components like antennas and capacitors is possible. Nano Dimension's [DragonFly IV](#) system can print a variety of metals and dielectrics with lateral resolutions of ~18 um on substrates of 16 x 16 cm$^2$. These tools cost hundreds of thousands of dollars rather than hundreds of millions.

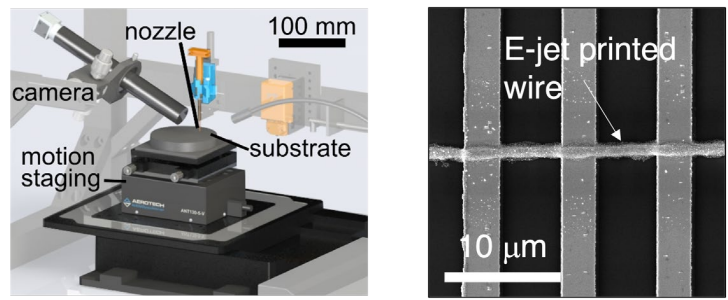

Figure 6: Schematic of a desktop-sized e-jet printing setup and a conductive metal line (horizontal) e-jet printed across photolithographically-patterned lines (vertical) courtesy of the Barton Lab at the University of Michigan.

4. Bottom-up processes have created modular proto-transistors

Components grown in bulk are on the cusp of feasibility. The Filler and Vogel Labs at Georgia Tech have created a "proto-transistor" via a process that can scale like a chemical reaction: they grew a semiconductor source, channel, and drain regions as well as a gate oxide aligned to the channel on a silicon nanowire [25]. While these proto-transistors need a metal-on-top-of-the-gate oxide to create the metal-oxide-semiconductor capacitor stack of functional transistors, the team established that the same method could produce that capacitor stack on wafers. The resulting capacitor performed similarly to those fabricated with photolithography [26].

In the limit, grown transistors could have comparable performance to etched silicon because of being made from a single crystal with a nanoscale channel length. Similar bulk methods could fabricate diodes, sensors, solar cells, or other components.

By integrating more materials and further miniaturization, the performance of nanomodular components could increase over time: high mobility materials like SiGe or InP can hit frequencies of several hundred gigahertz and it's possible to create transistors with gates as small as 1 nm [27].

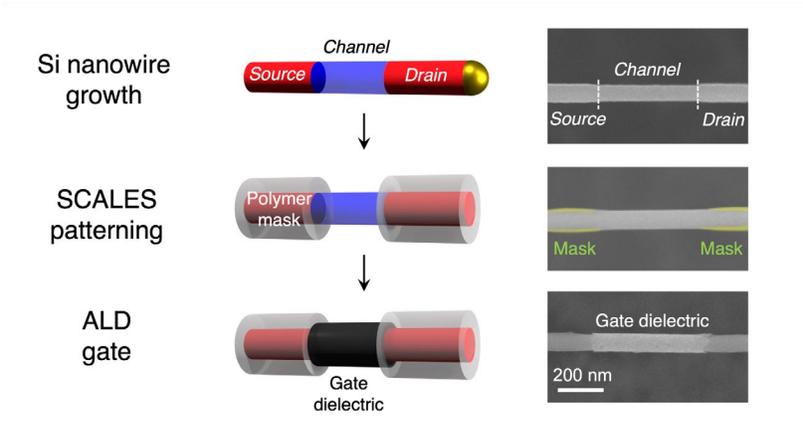

**Figure 7: Proto-transistor fabricated at Georgia Tech.**

## 5. Methods are emerging to dramatically scale up bottom-up component manufacturing

Geode Processing is a promising way to scale bottom-up component manufacturing [28]. The process grows single crystal semiconductor nanowires on the interior surface of powder microcapsules, like crystals in a geode. The microcapsule powder's high surface area coupled with the nanowire's growth rate indicates that hundreds of quintillions of transistors could be produced daily in a one cubic meter reactor – about as many transistors as the global semiconductor industry makes in a year [29]. That manufacturing scale would reduce component cost by orders of magnitude.

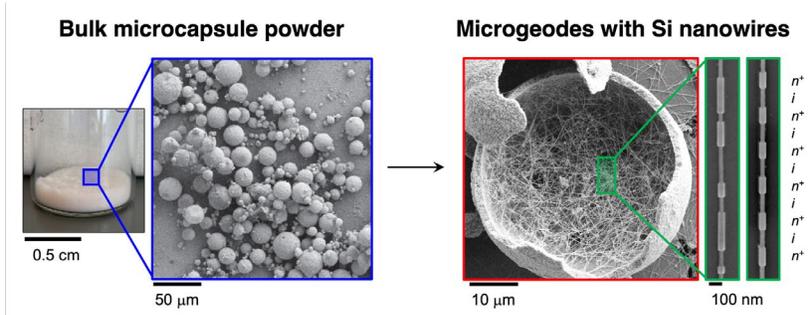

**Figure 8: The Geode Process enables scale-up of single-crystal semiconductor nanowires containing nanoscale segments by orders-of-magnitude.**

## 6. Traditional electronics manufacturing can bootstrap nanomodular electronics

Laboratories have demonstrated the steps needed to transform transistors manufactured via planar processing into modular components individually or in pairs. The steps include:

- Isolation and etching to permit lift-out,
- Contact prefabrication so wiring is easy during circuit creation,
- Fashioning degradation-proof packaging,
- Minor structural adjustments to properly sit components on the circuit substrate.

MicroLED technology [30] establishes that semiconductor components can be transferred from one surface to another with negligible loss of function. Small III-V-based light emitting diode pixels are first fabricated on conventional wafers before being transferred to a display backplane to be integrated with circuitry to drive each pixel. Current microLEDs measure 10-100 um per side with sub-10 um dimensions on the horizon.

Additionally, advances in semiconductor fabrication have made embedding components in a wafer unnecessary. While embedding was core benefit of the original planar process, state-of-the-art transistors like finFETs or nanowire FETs sit on top of the wafer and still achieve low voltages and high frequencies [31].

Laboratory work has explored the fabrication and mechanical transfer of other types of microscale and nanoscale devices. Transistors can be partially or primarily fabricated on-wafer and then "dry transferred" – moved without dispersion in liquid – to another surface with negligible performance loss [32, 33]. Researchers at PARC (Xerox) have demonstrated a suite of photocopier-like methods for transferring and deterministically arranging microscale chips, also without loss of function [34, 35].

## 7. Factoring processes into multiple steps enables new manufacturing paradigms

Some of the most transformational manufacturing innovations involve factoring an existing process step into multiple process steps using modular components as intermediates [1]:

- **Text manufacturing**. In the age of scriptoria, hand-copying of text was laborious, costly, and tightly controlled. Moveable type and the printing press factored text manufacturing into two distinct process steps: type fabrication and text fabrication. The resulting scalability of the new process changed the world.

- **Photography.** George Eastman factored the photography process into multiple steps. Film manufacturing and development were done by a specialist while the intermediate image collection step could be done by anyone using the Kodak camera.

The list goes on: DNA sequencing, continuous papermaking, and even the steam engine resulted from factoring existing processes.

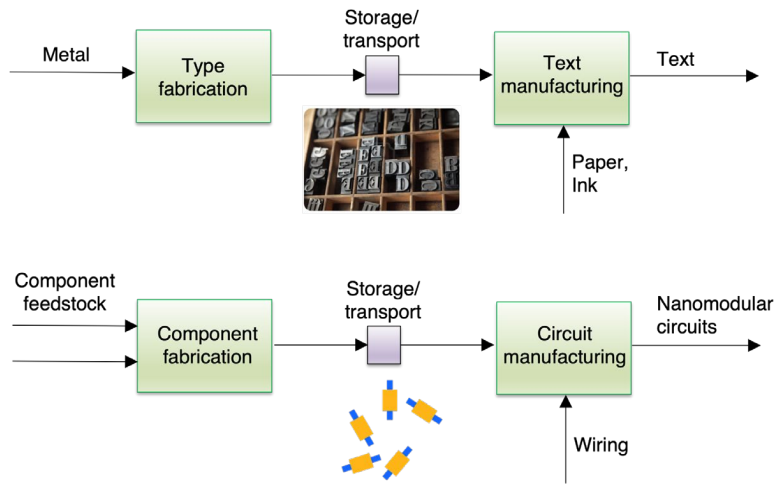

**Figure 9: Comparison of factoring in text and nanomodular electronics manufacturing.**

**Conclusion:** Specific technical evidence and historical analogy, when coupled with the challenges facing traditional microelectronics manufacturing, lead to one conclusion: nanomodular electronics must be tested.

# IV. Who cares? If you are successful, what difference will it make?

The ultimate goal of the program is to unlock the immense design space outside of traditional microelectronics. However, in the short term, there are at least two broad groups of people who should care about nanomodular electronics' success. Those who want to:

1. Implement algorithms in hardware, especially for **cryptography** and in edge **AI applications**.

2. Create **custom electronics on demand**, especially prototypers and groups with critical electronics needs that are vulnerable to supply chain disruptions.

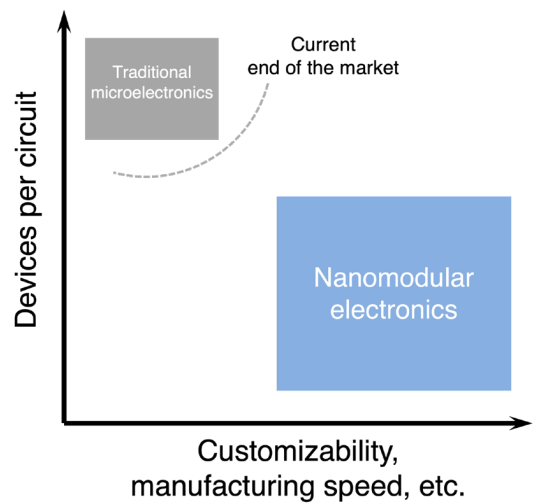

Figure 10: Comparison of the design space occupied by traditional microelectronics and the much larger opportunity presented by nanomodular electronics.

## A. Changing the rules for implementing algorithms in hardware

### 1. Cryptography

**Preventing hardware tampering.** Tampering with microelectronics at a hardware level enables malicious actors to steal data, render electronics inoperable, and potentially take control of entire electronic systems. Unfortunately, it is hard to detect or thwart.

By enabling organizations to fabricate their own microelectronics on their own tools, nanomodular electronics could drastically reduce the attack surface for tampering. Making it easier to create custom circuits would also enable organizations to move more algorithms off general-purpose devices by encoding them in hardware, reducing vulnerability to software hacks. (This security would initially come at a significant performance cost, limiting nanomodular electronics to truly mission critical work.)

**Speeding up physical cryptography.** Nanomodular electronics could improve security by enabling individual electronic parts to have unique cryptographic signatures. Analogous to how digital cryptography verifies digital identity, these signatures could generate, for example, a frequency response that verifies the identity of the hardware when combined with a "public key" [36]. Cryptographic signatures could enable organizations to secure, validate, and track high value parts, biologics, or other critical items or shipments.

Creating unique hardware signatures requires each circuit in a series to be different. While it's feasible for a single mask to contain many unique circuits, the maximum mask size as well as the cost and time to make it limits the scalability of this approach. However, creating a large run of unique circuits would be easy with nanomodular electronics.

## 2. AI on the edge

Nanomodular electronics would enable companies and individuals to implement artificial neural networks in hardware quickly and cheaply. These networks, the core of modern AI algorithms, consist of millions or billions of nodes with "weights" on the connections among them dictating how much one node affects another. Running neural networks general-purpose hardware like GPUs is slow and energy inefficient compared to custom hardware designed to perform specific tensor calculations. Hardware optimized for the specific neural network performs even better.

Speed and inefficiency are merely annoying when latency is unimportant and there's a good internet connection in AI applications like image generation or data analysis. However, latency and efficiency can make or break applications that depend on short response times, batteries, or bad internet connections like drone navigation and remote sensors. The ability to create a chip that hardwires a network can drastically speed up inference while reducing power consumption but is cost- and time-prohibitive except at the largest scales with current microelectronics manufacturing.

Nanomodular electronics are well-suited to neural networks because they are relatively robust to dropped connections and noisy computations. In addition to being a good fit for the more stochastic circuits in nanomodular electronics, it's possible to imagine eliminating the use of transistors for computation altogether. Limiting transistors to control might allow for much smaller, cheaper, more power efficient neural network implementations. Nanomodular electronics would create new opportunities for deploying neural networks without consistent power or internet connections (often referred to as "edge computing"), improve existing networks, and drastically speed up iteration cycles for both.

## B. Creating mission-critical custom electronics on-demand

### 1. Prototyping

**Enabling rapid prototyping**. Prototypers could use nanomodular electronics in models that are small, low-power, or highly customized. Iteration speed is essential to innovation: the ability to print custom circuits in-house could decrease development cycles bottlenecked by electronics: shortening them from days or weeks to hours. In the long run, the same electronics would be suitable for final products.

**Enabling more innovations to leave the laboratory.** Many research advances face a chicken and egg problem: an innovation needs to be incredibly valuable to warrant integration into *existing* processes, but it's hard to demonstrate the value of an innovation without getting it into the world. Tens of billions of dollars in public and private dollars fund work that sits on a shelf.

Nanomodular electronics could create a parallel system to incorporate new components more easily, improving returns on investment and increasing the diversity of electronics capabilities.

### 2. Decoupling from global supply chains

**Resilience to supply chain disruptions.** The COVID-19 pandemic disrupted the microelectronics supply chain, creating delays that are still being felt in 2022. Nanomodular electronics could give users a new option: the ability to make circuits in-house or locally.

In the long run, the large number of uses that do not require leading-edge performance could shift entirely to nanomodular electronics.

**Enabling new supply chains and ecosystems.** Factoring component fabrication and circuit creation into distinct steps relaxes the required know-how in any given firm. Manufacturers can specialize in fabricating a specific component or circuit type. The ability to specialize in that way would not only lower barriers to entry but could catalyze the emergence of new component and circuit-level supply chains and ecosystems. Such shifts could create opportunities for more electronics to be made within single countries like the United States.

**Eliminating shipping delays.** Despite the speed and extent of today's global supply chain, there are many situations or locations for which it is inadequate: communication is cut off in the weeks it takes a cell phone chip to arrive in a rural village; tens of thousands of dollars in revenue are lost in the day it takes for a chip to repair a commercial airplane's circuit board to arrive.

Nanomodular electronics could minimize or eliminate the problems of shipping delays by allowing the means of production – the printer – to be in many more locations.

## C. A vision for nanomolecular electronics

On a long timeline, nanomodular electronics could make transistor manufacturing resemble chemical manufacturing and make electronics resemble software. Chemicals are produced in bulk processes that scale with volume. While transistors are remarkably cheap compared to the past, they are orders of magnitude more expensive on a per unit basis than even the most expensive drugs. Creating transistors the same way we manufacture chemicals could change the role of electronics in the world, making them a truly ubiquitous commodity.

Software is [composable](), transferable, and easily distributed. (One can combine different pieces of software that know nothing about each other. Assuming the source code is available, one can modify a piece of software for unique purposes. Anybody with a computer can write their own software and even innovate on how software is made: anybody can create and distribute modifications of the Linux source code, for example.) If electronics had these properties, the structure of the industry would transform. Instead of the consolidation that exists today, thousands of small electronics shops and startups could emerge, constantly innovating. This restructured industry could in turn create a Cambrian explosion in how we use electronics in the same way that personal computers led to computers being used in ways that Von Neumann and other pioneers could never have imagined. It is possible to imagine obscenely efficient edge computing, truly disposable electronics, electronics seamlessly integrated into biological systems, and ubiquitous physical cryptography.

The chip-related supply chain issues that plague downstream industries from cars to air conditioners [37] would be a thing of the past. Mass customization could make it much easier to repair the ever-increasing number of devices that depend on electronics, empowering consumers. Varied geographies could specialize in specific types of microelectronics production, stimulating local economies and letting Smithian growth through specialization do its magic [38].

Electronics would not just be manufactured in new places around the world, but off of it as well. The potentially small footprint of a nanomodular electronics printer and low expertise requirements mean that astronauts could manufacture microelectronics in space or on another planet. This capability in turn could support a burgeoning space manufacturing industry and make space exploration more robust to equipment failures.

Finally, it is worth noting that the complexity of modern electronics manufacturing is a key bottleneck to creating fully self-replicating factories. Speculatively, nanomodular electronics could unlock a fully autonomous economy in which factories that require nothing but robotically-supplied raw materials could create more factories that could then produce goods close to where they are used, creating unprecedented abundance around the world and beyond it.

# V. What are the risks?

## A. Risks to the success of a 5-year program

While the biggest risks to research are the "unknown unknowns," precisely describing the "known unknowns" helps prioritize work [39]. This section describes the biggest known technical risks to nanomodular electronics on the timescale of a five year program:

1. Suspending transistors in an ink may make it hard to wire them together.

2. Components are susceptible to contamination.

3. Adaptive routing may not scale.

This section describes these risks and others, while Section VI maps out "derisking" projects designed to address them early in the program.

**Suspending transistors in an ink may make it hard to wire them together.** Storing transistors in a colloidal solution ("ink") makes them transportable and printable. The van der Waals forces among nanoscale objects tends to result in irreversible clumping without special techniques. While these approaches have worked on simple nanoscale structures, a transistor's complex combination of doped semiconductors, oxides, and metals may cause problems. A coating could prevent components from clumping but it would need to be removed before component wiring, adding unwanted process complexity.

Demonstrating methods to store and transport modular components in a colloidal solution that still permits wiring is a key part of the derisking program.

**Components are susceptible to contamination.** Placing components in a liquid increases the risk of contamination by performance-degrading impurities like metal ions. Impurities in the gate stack of a transistor, for example, can decrease carrier mobility, shift the turn-on voltage, and more.

The risk of contamination stems from multiple factors. Liquids tend to solvate impurities, effectively concentrating them near components. The exposed contacts of modular components, while making it easy to wire components together, could serve as an entry point for impurities into the component's interior. Once inside, the small size of the components means that impurities could quickly reach any sensitive structure. (Impurity diffusion rates increase with the inverse square of distance: the time required for impurity diffusion over 10 nm is $10^6$ times shorter than diffusion over 10 μm.) Once contaminated, it would be hard to remove impurities without degrading the component in other ways.

Packaging components before they are put into an ink could prevent contamination, but it also has the potential to complicate the wiring process.

Establishing methods to prevent excessive component contamination and performance degradation is also a thrust of the derisking effort.

**Adaptive routing may not scale.** Nanomodular electronics faces sensing challenges:

- How to detect the position and orientation of imperfectly placed components.
- How to dynamically transition from a high-level circuit design to a specific circuit layout.
- How to create redundant circuits that work with defective components.
- How to generate multiple layers of components.

The biggest risk is that adaptive routing runs into "the curse of dimensionality" [40] as it scales to thousands or millions of components.

As the scaling of adaptive circuit routing will determine nanomodular electronics' usefulness, it is another focus of the derisking strategy.

## B. Risks beyond the timescale of the program

The ultimate goal of this nanomodular electronics program is to change how electronics are made and put a powerful new tool in humanity's toolbelt. There are many ways the technology could fall short of that goal, even if the five-year program achieves all its objectives.

### 1. The technology may not be adopted

Success is not fully realized with a "working" nanomodular electronics manufacturing system. Additional development would happen in one of four ways (and potentially a combination thereof):

- Amateurs tinker the technology towards adoption as open source software, 3D printers, and personal computers evidence.
- A large company makes aspects of the technology its own, like the trajectories of vacuum tubes and many chemical processes.
- A startup identifies a niche and uses the technology as a wedge to enter more markets, like mRNA vaccines.

- The government pays for a functionality that the market would not otherwise fund as was done with transistors and GPS receivers.

While the continuous involvement of interested and knowledgeable amateurs carries many advantages (see [Section I](#)), the system created for the competition could be too limited or frustrating for serious engagement. Student competitions also fail for reasons unrelated to the technology – being poorly run, perceived as not worth the time or effort, insufficient funding, and the like.

## 2. The technology may become niche

Nanomodular electronics could be viable for a niche application and a small group of users. Examples of this failure mode include e-ink, lighter-than-air aircraft, and (so far) augmented- and virtual-reality headsets. Nanomodular electronics may be useful for adding cryptographic security to electronics at the hardware level and get "stuck" there.

This program mitigates that risk by targeting a student competition to force technology generalization. While a gambit – increasing the risk that the technology will not be useful for anything to avoid becoming trapped in a niche – it provides the most promising approach.

## 3. The technology may run into long-term technical risks

Upon creating a working system, technical limitations could prevent it from reaching GPT status. Several technical risks are:

- *Circuit size:* Nanomodular circuits could hit an upper limit on how many transistors and other components a functional circuit may contain before its performance degrades.

- *Circuit yield:* The upper limit on circuit yield may be low, even with robust wiring schemes, redundancy, and methods to remove and replace defective components or wires.

- *Lifetime:* There's a chance that nanomodular electronics will have shorter lifetimes than traditionally manufactured electronics.

None of these long-term technical risks make nanomodular electronics unusable, they merely limit the applications.

## 4. Competing standards for nanomodular electronics might make interoperability hard

Upon adoption, multiple standards could hamstring the upsides of modularity. Nanomodular electronics with heterogeneous components from many sources and materials need to work together. Multiple competing standards would reduce the range of applications and fracture further development efforts.

# VI. How is the program structured? How long will it take?

The program is setup in three phases:

A. **Derisking projects** to show that there are ways to mitigate the biggest known existential risks to nanomodular electronics.

B. **Proof-of-concept** projects to build minimally-viable components of the system.

C. **System integration** projects to develop a minimal viable system and then iterate on that system to hit the program's performance goals.

The derisking projects will require **approximately one year**. Proof-of-concept projects and system integration should require **an additional 4 years**.

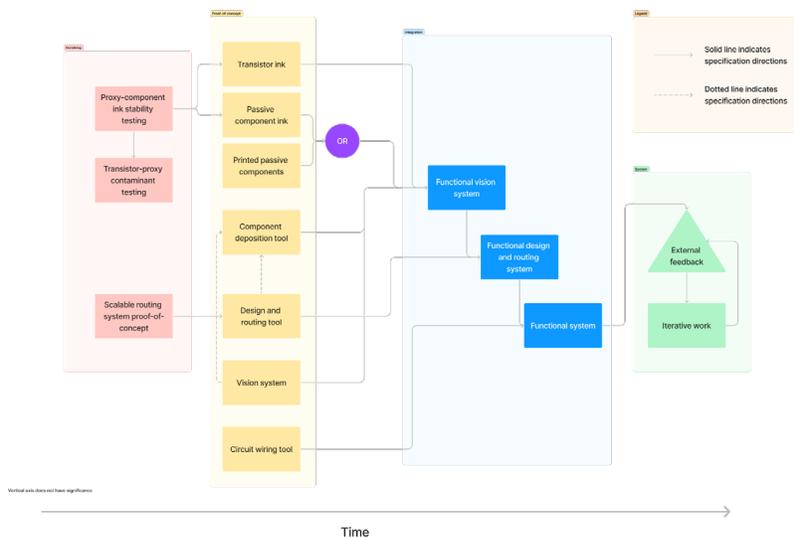

**Figure 11: A dependency diagram for the work in this program to give a sense of potential paths. Link to a detailed living document.**

The discussion in this section assumes that the program has enough resources to run all of a phase's projects in parallel. Less money will increase timelines and more money may decrease them.

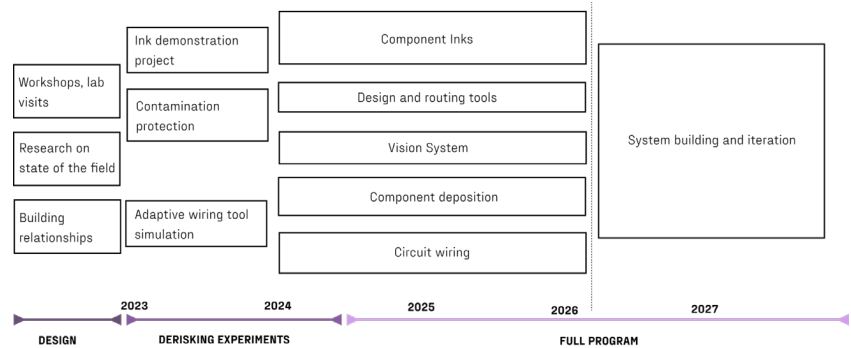

**Figure 12: Overview of program structure.**

## A. Derisking projects

The three projects to address the major risks to nanomodular electronics (see Section V) are:

1. **Ink viability**: Making a functional ink that can simultaneously keep semiconductor components from bunching without ruining their ability to connect to wires or introducing contamination.

2. **Contamination**: Showing that components do not accumulate enough contaminants to degrade their performance.

3. **Circuit routing**: Showing that an adaptive routing system can handle large numbers of components.

These projects set the stage and ensure consistent prioritization for the remainder of the program.

### 1. Ink viability

Ink(s) must:

- Maintain separation among and between components.

- Not require excessive processing before component wiring.

- Not create excessive contact resistance during component wiring.

- Not introduce contaminants that undermine component performance (which will be tested in project A2).

The project will need proxy silicon components. While exact implementation is in the hands of the lab doing the project, it's important that the chosen proxy components successfully capture the relevant information about the ink separation and contacting. The components suspended in solution don't necessarily need to be the same as the components that test for contact resistance or contamination, although it would be a more convincing demonstration.

Two possible proxies for testing separation and contact resistance are either resistors fabricated on a silicon substrate and then lifted off or semiconductor or semiconductor particles with analogous size and surface properties to future functional components.

The project will be successful if more than 90% of silicon proxies with sub-micron dimensions can remain unaggregated in the ink for 24 hours and yield a contact resistance below 1 $\mu\Omega$-cm$^2$ upon wiring at temperatures below 200 °C in air with acceptable levels of contamination (as determined by project A2.)

## 2. Contamination

The goals for this project are:

- Creating a proxy that accumulates contaminants in the same way as would a transistor in a nanomodular electronics system.

- Measuring the effects of atmospheric conditions and ink composition on component contamination levels and thus performance.

The contaminant proxy does not need to be dispersed into solution if its exposed surfaces interact with the solution in the same way as a fully suspended component. Ideally, the proxy would be easy to produce and make measurement easy. The specific contamination measurements will depend on the proxy.

Unpackaged transistors are one possible proxy that would meet these requirements, but there may be others. For a transistor-based proxy, the key measurements would be threshold voltage, subthreshold swing, effective mobility, and interface state density.

To some extent, derisking projects A1 and A2 are a generator/discriminator pair: project A1 creates inks and project A2 tests their viability. The separation acts as an incentive for both groups: the performers on project A1 will want to ensure the proxies are accurate test subjects and the performers on project A2 will want to give realistic assessments. The relationship should be collaborative, not adversarial: ideally, the project A2 testers will give feedback to A1 participants on ways to improve the solution.

For a transistor-based proxy, the project would be successful if it can maintain a threshold voltage near 1 V, a subthreshold swing below 100 mV/decade, an effective mobility above 300 $cm^2$/V-s, and device-to-device variations in these properties less than 10% after exposure to both an ink solution and air for 24 hours.

### 3. Circuit routing

The three goals for this project are:

- Creating a system for simulating stochastically deposited components and creating connections among them to match a circuit diagram.

- Understanding the necessary control of component placement, if any, and how this depends on component number.

- Finding the scaling limits of adaptive routing.

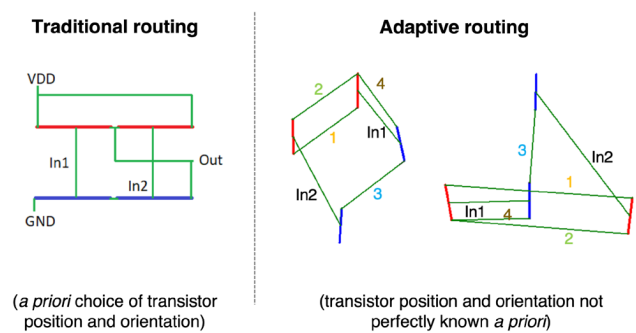

**Figure 13: Comparison of traditional circuit routing and routing that compensates for component placement error.**

One approach to this project would be to build a routing tool and scale it to large stochastic circuits. Another approach would be to show that there are theoretical scaling limits that apply to any system or that there should be no limits without building a simulation.

This project will be successful if it can demonstrate interconnections for a set of test architectures with 10,000 simulated components with a positional variance of at least ±2 um and an orientation variance of at least ±20°.

## B. Proof-of-concept projects

There are five system pieces needing proof-of-concept work:

1. Component inks
2. Design tools
3. Vision system
4. Component deposition
5. Circuit wiring

Some proof-of-concept projects explore different process trade-offs, particularly between precision and complexity in process steps and tooling. These results will point towards the most promising trajectories for Phase III projects and further development.

The proof-of-concept work should take 24 months. The most likely bottlenecks are the creation of minimal component inks in project B1 for projects B4 and B5 to use (12-18 months) and the wiring of components in project B5 (another 12-18 months).

### 1. Component inks

The goal of this project is to bootstrap nanomodular components from traditional semiconductor manufacturing and support the other projects.

This project will produce transistor and resistor inks based on the results from project A1. While transistors are essential components, resistors have fewer fabrication steps, their function is less sensitive to structural and chemical details, and other proof-of-concept projects can use them.

To facilitate knowledge transfer to components created without silicon wafers or photolithography, the component structures should resemble those possible with bottom-up methods – the dimensions and shapes like those of grown nanowires.

The key metric for these components is the ability to match the voltages and frequencies of 180 nm node processes from the late 1990s. Recall that creating circuits with similar performance to 1990s-level processes is the eventual goal of the entire program.

|  | nMOS | pMOS |
|---|---|---|
| Supply voltage | 1.3-1.5V | 1.3-1.5V |
| Oxide thickness | 3 nm | 3 nm |
| $L_{gate}$ | 130 nm | 150 nm |
| $V_T$ | 0.3 V (130 nm) | -0.24 V (150 nm) |
| $I_{Dsat}$ (1.5 V) | 0.94 mA/μm | 0.42 mA/μm |
| $I_{off}$ | 3 nA/μm | 3 nA/μm |
| $C_J$ (0 V) | 0.65 fF/μm$^2$ | 0.95 fF/μm$^2$ |
| Silicide | 3-5 Ω/sq | 3-5 Ω/sq |

**Figure 14: Table of transistor parameters for the 180 nm process node [2].**

This project will be considered successful if it creates transistors and resistors that function after being suspended in the solution identified in project A1, deposited by the tools from project B4, and wired together by project B5. The active components need to have a threshold voltage less than 1 V and frequencies greater than 1 GHz. More than 90% of the components should remain unaggregated after 24 hrs.

Good jumping off points for this project include this work on semiconductor device fabrication and transfer as well as this work on colloids.

The rough estimate is that creating a resistor and then transistor ink would each take 2 people and 12-18 months to complete.

2. Design tools

This project will create software for routing interconnections between imperfectly placed components to create specified circuits. Ideally, the tool will create circuits from randomly positioned and oriented components. The ability to handle arbitrary component placement would relax constraints on component deposition (project B4) and make position and orientation control all upside. However, the project can still be successful if the tool can handle a scoped amount of position and orientation randomness.

The software needs to handle enough different components to build "interesting" circuits. Additionally, it needs to deal with some component variability and imperfect wiring.

There are at least 2 approaches to building the requisite tool:

**Modifying existing tools.** The design tool could modify existing tools to handle imperfect component placement and strip out unnecessary complexity for initial applications. The advantage of this approach is that tools like [OpenROAD](#) have hundreds of developer hours and can already handle digital, analog, and mixed-signal circuits. The potential downside is that existing tools encode many abstractions that assume the constraints of traditional electronics manufacturing, and working within those frameworks might be more trouble than it's worth. [This work](#) is a good example of combining existing design tools with custom point tools.

**Built-from-scratch.** A tool built from scratch could leverage a limited number of components, variability in component performance, and imperfections in component placement to create a framework specifically for nanomodular electronics. Instead of offering the complete design control of off-the-shelf tools, this approach could focus on design abstractions that are easier to use for nanomodular electronics and only be as complex as the Nanomodular Electronics Challenge demands.

While initial design tool creation does not block nor is it blocked by any other project, it is coupled to component fabrication (B1), component deposition (B4), and wire creation (B5). The amount of stochasticity the software can handle and still create performant circuits (in simulation) will determine the precision requirements of component deposition.

Conversely, the precision and size of the interconnect creation will determine the performance needs of the design tool.

This tool will be successful if it can handle components with a density of 1 component per 100 square microns and a maximum position/orientational variance set by the adaptive routing mitigation project (A3). The tool needs to create circuits with at least 5,000 of the 10,000 components identified by the vision system (B3).

Ideally, both approaches can be done in parallel. Each approach should take 2 people and 12-18 months to complete.

### 3. Vision system

This project will create a vision system able to detect the position and orientation of stochastically deposited components to enable the routing tools to find the wiring paths to make specified circuits.

Ideally, the vision system won't place any demands on the components, but it may need to tag components with something that emits or reflects light to achieve high enough resolutions. Needing to instrument the components is acceptable as long as it does not interfere with component performance in project B1 and works after being put into the inks from project A1.

This project will be successful if the vision system can find the position and orientation of at least 10,000 components in a single plane to within an accuracy of half the component critical dimension (*e.g.,* the channel length of a transistor). The real test is whether it works with the deposition method created in project B4.

The identification and contacting of nanowires in [this work](this work) is a good starting place for this project.

This project should take 2 people 12 months to complete.

### 4. Component deposition

This project will create a printer that deposits nanomodular components using the inks generated in project A1. The component deposition project has several dimensions:

- Controlling the position and movement of the deposition nozzle

- Component deposition
- Control systems
- Substrate design

The project's goal is to create arrays of components with a desired density rather than precise placement. Although precision placement would enable more complex circuits, arrays of the same component are sufficient for the initial competition.

There are at least 2 broad approaches to depositing components:

**Precise placement with complex tooling and processes.** This approach assumes placement precision is critical, akin to what is possible with traditional microelectronics. Better than random placement of components might be limited to a single substrate or component type. There are many possible approaches including chemically, electrically, magnetically, flow, or optically controlled deposition. These papers on nano and micro-object assembly are jumping off points for the high-resolution assembly sub-project.

**Average placement with simple tooling and processes**. This approach places a premium on tooling simplicity and versatility rather than placement precision. Deposition via spray coating or blade coating, for example, could be possible on a range of substrates/surfaces (*e.g.,* glass, plastic, paper), giving Nanomodular Electronics Challenge competitors an extra degree of design freedom. Adding new component types would also be easier. This work on nanowire spray coating and capillary printing are good points of departure for the low resolution assembly sub-project.

This project will depend on the proxy inks developed during project A1 so that work can proceed without the functional inks from project B1. A feedback loop between the ink project (B1) and this deposition project is important because of the coupling between ink properties and deposition mechanisms. The precision and accuracy achieved by the deposition project will drive the design tool project (B3) as well as the circuit wiring project (B5). Specifically, the precision and accuracy of component placement sets the level of stochasticity that the design tool needs to accommodate as well as the precision and turning radius of the wires.

This project will be successful if the deposition tool can deposit components from 2 inks in a single layer with an average density less than 1 component per 100 square microns, across an area of at least 0.5 cm$^2$, and a maximum position/orientational variance set by the design tool (B2). It is important to keep the number of touching or overlapping components small, but the vision system (B3) will be able to identify and route appropriately.

Ideally, several approaches can be done in parallel. Each approach should take 2 people 12-18 months to complete.

### 5. Circuit wiring

This project will build a tool that lays down conductive wiring among deposited active components and deposits insulating material as needed to isolate wires from the environment and each other. The project needs to determine materials and chemistry for forming the connections, characterize wire properties, and create the tool that prints them. While the default approach is to create colloidal inks of passive components, a stretch goal for the project would be to test the viability of directly printing passive components instead [41].

This project can use the proxies developed during projects A1 and A2 so that work progresses along with the component ink (B1) and deposition (B4) projects.

Important wiring properties are:

- Conductivity (low conductivity makes circuits slower and heat up faster)

- Width (smaller wires can connect to small components)

- Contact resistance (high resistance connections degrade performance)

- Printing conditions (no overly special equipment)

The tooling needs to print wires quickly enough to turn around circuits for a competition, be small enough to ship, and be reliable enough to maintain circuit yield. This work on e-jet printing and this work on fountain pen nanolithography are good reference points for the wire printing project – both approaches could potentially hit the project's success metrics.

For the project to be successful, the wires contacting the component should be no greater than 150 nm wide, with conductivity greater than $10^5$ /Ω-cm and contact resistance less than 1 $\mu$Ω-cm$^2$. The wires and contacts should be created at temperatures less than 200 °C in air. Wiring among components can be as wide as 1 $\mu$m.

The tool that creates the wires needs to work at a rate greater than 1 mm/sec. Wire shorts need to be in the range of 1 per 10,000 wires. Minimizing fail opens is also important but easier to handle with circuit design (*e.g.,* including redundant components). Insulators must have a relative dielectric constant less than 3.9 (*i.e.,* that of silica).

A rough estimate is that it could take 2 people 12-18 months to complete each approach.

## C. System integration

System integration has two major elements:

- Developing a single system from the components developed in phase 2.

- Improving the components that bottleneck the system's performance.

In addition to integrating the outputs of the proof-of-concept projects, a key systems-level challenge will be switching from passive stub electronic components to transistors. Although the proof-of-concept projects will demonstrate the ability to deposit active components, the initial integration will use only passive components. This approach decouples building working active components from a working deposition+wiring system.

Depending on results from different component-level projects, the projects to address bottlenecks may include:

- Shrinking components

- Improving wire resolution to handle smaller critical dimensions

- Making larger quantities of components

- Improving placement precision

- Decreasing contact resistance

In parallel, the program will be working on components that do not depend on the traditional semiconductor manufacturing process, such as fully bottom-up transistors. However, development of a performance level where they work with the rest of the system is beyond the scope of this program without a great deal of serendipity.

We expect this phase of the program to take 24 months – limited first by the work to create a single system (which we expect to take 12 months) and then by the speed at which teams can iterate to bring that system's performance to the point where it works well enough to support the Nanomodular Electronics Challenge.

# VII. What are the milestones for each phase of the program?

## A. Success metrics for risk mitigation projects

| Description | Quantity | Unit | Project |
|---|---|---|---|
| Maximum silicon proxy dimension | $1 \times 10^{-7}$ | m | A1 |
| Minimum fraction of proxies remaining unaggregated after 24 hrs | 0.9 | | A1 |
| Maximum contact resistance | 1 | $\mu\Omega\text{-cm}^2$ | A1 |
| Maximum wiring temperature | 200 | °C | A1 |
| Maximum component threshold voltage | 1 | V | A2 |
| Maximum component subthreshold swing | 100 | mv/decade | A2 |
| Minimum component effective mobility | 300 | $cm^2/V\text{-s}$ | A2 |
| Maximum component-to-component variation | 10 | % | A2 |
| Minimum number of circuit components handled by the interconnect system | 10,000 | | A3 |
| Minimum component position variance handled by the interconnect system | ± 2 | $\mu m$ | A3 |

| Minimum component orientation variance handled by the interconnect system | ± 20 | ° | A3 |

## B. Completeness checks for component-level projects

| Description | Quantity | Unit | Project |
| --- | --- | --- | --- |
| Maximum component threshold voltage | 1 | V | B1 |
| Minimum component frequency | 1 | GHz | B1 |
| Minimum number of components handled by the design tool | 10,000 | | B2 |
| Minimum number of components wired per circuit | 5,000 | | B2 |
| Maximum position variance handled by the design tool | Set by project A3 | | B2 |
| Maximum orientation variance handled by the design tool | Set by project A3 | | B2 |
| Minimum number of components vision system can identify | 10,000 | | B3 |
| Maximum error on recognized component position | 50% comp. critical dimension | | B3 |
| Maximum error on recognized component orientation | ± 15 | ° | B3 |
| Minimum number of nanomodular component inks deposition tool can handle at once | 2 | | B4 |
| Minimum component areal density | 0.01 | $\mu m^{-2}$ | B4 |
| Minimum component deposition area | 0.5 | $cm^2$ | B4 |
| Maximum component position variance handled by deposition tool | Set by project B2 | | B4 |
| Maximum component orientation variance handled by deposition tool | Set by project B2 | | B4 |
| Minimum wire width for contacting components | 150 | nm | B5 |

| Maximum wire width for interconnecting components | 1 | $\mu$m | B5 |
| --- | --- | --- | --- |
| Minimum conductivity | $10^5$ | /$\Omega$-cm | B5 |
| Maximum contact resistance | 1 | 1 $\mu\Omega$-cm$^2$ | B5 |
| Maximum wire printing temperature | 200 | °C | B5 |
| Wiring atmosphere | Air | | B5 |
| Minimum wire print rate | 1 | mm/sec | B5 |
| Percentage shorted wires | 0.01 | % | B5 |
| Maximum insulator relative dielectric constant | 3.9 | | B5 |

## C. Final checks for competition focused program

| Description | Quantity | Unit |
| --- | --- | --- |
| Minimum frequency for a created flip-flop circuit | 100 | MHz |
| Minimum frequency for a created differential circuit | 100 | MHz |
| Maximum circuit footprint for flip-flop and differential circuit | 0.05 | mm$^2$ |
| Maximum time to create flip-flop and differential circuit | 10 | min |
| Maximum design tool failure rate | 0.1 | /circuit |
| Maximum printer failure rate | 0.5 | /hour |

These metrics are a proxy for the real goal of the program, which is to create a system that can support a competition. The acid test will be feedback from the actual teams using it.

# Acknowledgments

Michael Filler and Eric Vogel co-created the concept of nanomodular electronics. A nanomodular electronics research program or any future technology will exist in no small part due to Eric's insight, intellectual honesty, and pragmatism.

Many of the initial contexts of use for nanomodular electronics came from the enormously creative and generative participants of the 2022 Workshop on On-demand Integrated Circuits: Gregory Abowd, Kira Barton, Marc Berte, Asif Bhatti, Mohit Bhoite, David Brock, Ahmed Busnaina, Mario Cruz, Paul Franzon, Tim Hancock, Tina Kaarsberg, Alex Kozak, Matt Parlmer, Nadya Peek, Matthew Realff, Eric Vogel, Gang Qu. Thank you to Schmidt Futures and Convergent Research for their generous support of the event.

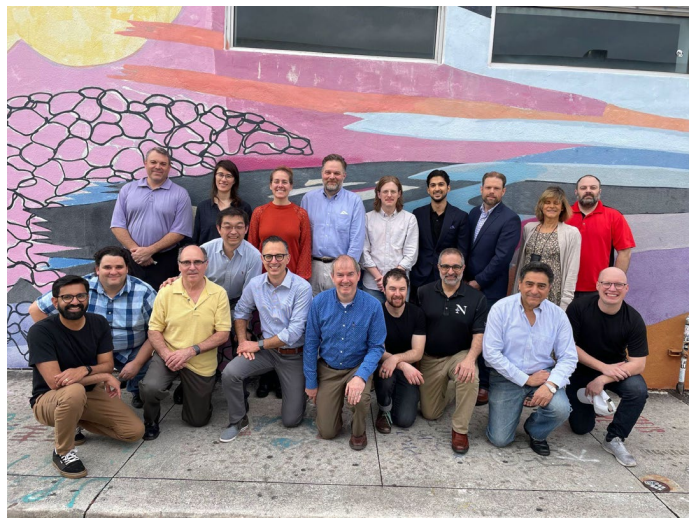

We are grateful to everyone who grappled with early drafts, helping to polish and hone our arguments: Eric Vogel, Matthew Realff, Sarah MacLeod, Tom Kalil, and Marc Berte.

# Appendix: Other potential program goals

### Build a common circuit using nanomodular electronics

A straightforward goal is to use a new process to create the same product as the old process. Building a common circuit forces the system to create results directly comparable to those of traditional microelectronics processes. Building a common circuit creates a clear definition of "success" that could provide paths for improvement once a working system exists. The power of clear metrics and an unambiguous definition of success is not to be underestimated.

The [555 timer](#), originally designed in 1970, is "probably the most popular integrated circuit ever made." While there are other circuits that are equally simple or well-characterized, reproducing an icon (by some estimates more than a billion 555 timers are created each year) in a new way is not insignificant. A 555 timer includes several diverse components, both analog and digital, making it a good target. This component diversity would ensure a nanomodular electronics system that could create a 555 timer could also create many other circuits.

The metrics that matter for a 555 timer are the circuit's physical volume and bandwidth, and the process build time, build volume, and number of substrate materials. The correct target values for these metrics need to be realistic, but aggressive enough to force real innovation. Attendees of the Workshop for On-Demand Integrated Circuits identified a 1 mm$^3$ circuit volume, 1 MHz bandwidth, built in 1 day, on two different materials, within a system volume of 1 m$^3$ as a promising target.

While creating a 555 timer would prove the viability of factoring component and circuit production, it does not help anybody do anything they want to do. It also neither forces nor excludes generality.

## Fabricate cryptographic circuits using nanomodular electronics

Building circuits with unique signatures that can indicate tampering is a promising early application of nanomodular electronics. Tamper-detection in chips is critical to national security: malicious actors introduce vulnerabilities into microelectronics at the hardware level and chips pass through many hands both during and after their manufacture. The program could aim to create a system to build tamper-detecting circuits and verify functionality with a red-teaming cryptography competition.

In this competition, several groups ("red teams") would attempt to tamper with circuits built during the program. Program participants would then attempt to determine which circuits had been affected. The program's "grade" would be determined based on the number of false negatives (tampered circuits that appeared uncompromised) and false positives (circuits that appeared to have been tampered with but were uncompromised). A perfect score would be zero false positives and negatives.

Developing physical cryptography has several advantages. Hardware-level cryptography is an obvious near-term application for nanomodular electronics. Demonstrating the capability early would generate a lot of interest. Many organizations (like the US military and large corporations) have significant resources, increasing the chance of further development. A concrete application provides a powerful "context of use," feedback that would increase the change of technology adoption and long-run impact.

Unfortunately, a cryptography goal would sideline development of a GPT. The system design requirements make it unique to the organization that develops it. There would be no need or desire to explore a system to support a wide array of applications, nor would it need to have tight specifications. While a cryptographic goal targets a serious context-of-use, the bar for cryptographic uses is much higher than simply enabling students to experiment. It is also unlikely that a single program would advance the technology enough to create a strong feedback loop with excited users.